\documentclass[aip,rsi,preprint,groupedaddress,showmsc,superscriptaddress,floatfix]{revtex4}
\usepackage{amsmath,amsfonts,amssymb,caption,hyperref,color,epsfig,graphics,graphicx,latexsym,mathrsfs,revsymb,theorem,url,verbatim,epstopdf}

\usepackage[center]{titlesec}
\titleformat{\section}[hang]{\Large\bfseries\filcenter}{}{1em}{}
\titleformat{\subsection}[hang]{\bfseries}{}{1em}{}
\usepackage{amsmath,amsfonts,amssymb,caption,hyperref,color,epsfig,graphics,graphicx,latexsym,mathrsfs,revsymb,theorem,url,verbatim,epstopdf,cleveref}
\hypersetup{colorlinks,linkcolor={blue},citecolor={blue},urlcolor={red}}

\newtheorem{definition}{Definition}

\newtheorem{lemma}[definition]{Lemma}

\newtheorem{theorem}[definition]{Theorem}
\newtheorem{corollary}[definition]{Corollary}

\def\squareforqed{\hbox{\rlap{$\sqcap$}$\sqcup$}}
\def\qed{\ifmmode\squareforqed\else{\unskip\nobreak\hfil
\penalty50\hskip1em\null\nobreak\hfil\squareforqed
\parfillskip=0pt\finalhyphendemerits=0\endgraf}\fi}
\def\endenv{\ifmmode\;\else{\unskip\nobreak\hfil
\penalty50\hskip1em\null\nobreak\hfil\;
\parfillskip=0pt\finalhyphendemerits=0\endgraf}\fi}
\newenvironment{proof}{\noindent \textbf{{Proof.~} }}{\qed}
\def\Dbar{\leavevmode\lower.6ex\hbox to 0pt
{\hskip-.23ex\accent"16\hss}D}
\def\bpf{\begin{proof}}
\def\epf{\end{proof}}
\newcommand{\abs}[1]{\left\lvert {#1} \right\rvert}

\newcommand{\nc}{\newcommand}

\def\bea{\begin{eqnarray}}
\def\eea{\end{eqnarray}}
\def\beq{\begin{equation}}
\def\eeq{\end{equation}}
\def\bal{\begin{aligned}}
\def\eal{\end{aligned}}
\def\bma{\begin{bmatrix}}
\def\ema{\end{bmatrix}}

\def\diag{\mathop{\rm diag}}

\def\a{\alpha}

\def\z{\zeta}

\def\t{\theta}

\def\p{\pi}

\def\o{\omega}

\nc{\bbA}{\mathbb{A}} \nc{\bbB}{\mathbb{B}} \nc{\bbC}{\mathbb{C}}
\nc{\bbD}{\mathbb{D}} \nc{\bbE}{\mathbb{E}} \nc{\bbF}{\mathbb{F}}
\nc{\bbG}{\mathbb{G}} \nc{\bbH}{\mathbb{H}} \nc{\bbI}{\mathbb{I}}
\nc{\bbJ}{\mathbb{J}} \nc{\bbK}{\mathbb{K}} \nc{\bbL}{\mathbb{L}}
\nc{\bbM}{\mathbb{M}} \nc{\bbN}{\mathbb{N}} \nc{\bbO}{\mathbb{O}}
\nc{\bbP}{\mathbb{P}} \nc{\bbQ}{\mathbb{Q}} \nc{\bbR}{\mathbb{R}}
\nc{\bbS}{\mathbb{S}} \nc{\bbT}{\mathbb{T}} \nc{\bbU}{\mathbb{U}}
\nc{\bbV}{\mathbb{V}} \nc{\bbW}{\mathbb{W}} \nc{\bbX}{\mathbb{X}}
\nc{\bbZ}{\mathbb{Z}}

\nc{\bA}{{\bf A}} \nc{\bB}{{\bf B}} \nc{\bC}{{\bf C}}
\nc{\bD}{{\bf D}} \nc{\bE}{{\bf E}} \nc{\bF}{{\bf F}}
\nc{\bG}{{\bf G}} \nc{\bH}{{\bf H}} \nc{\bI}{{\bf I}}
\nc{\bJ}{{\bf J}} \nc{\bK}{{\bf K}} \nc{\bL}{{\bf L}}
\nc{\bM}{{\bf M}} \nc{\bN}{{\bf N}} \nc{\bO}{{\bf O}}
\nc{\bP}{{\bf P}} \nc{\bQ}{{\bf Q}} \nc{\bR}{{\bf R}}
\nc{\bS}{{\bf S}} \nc{\bT}{{\bf T}} \nc{\bU}{{\bf U}}
\nc{\bV}{{\bf V}} \nc{\bW}{{\bf W}} \nc{\bX}{{\bf X}}
\nc{\bZ}{{\bf Z}}

\nc{\bmA}{{\bm A}} \nc{\bmB}{{\bm B}} \nc{\bmC}{{\bm C}}
\nc{\bmD}{{\bm D}} \nc{\bmE}{{\bm E}} \nc{\bmF}{{\bm F}}
\nc{\bmG}{{\bm G}} \nc{\bmH}{{\bm H}} \nc{\bmI}{{\bm I}}
\nc{\bmJ}{{\bm J}} \nc{\bmK}{{\bm K}} \nc{\bmL}{{\bm L}}
\nc{\bmM}{{\bm M}} \nc{\bmN}{{\bm N}} \nc{\bmO}{{\bm O}}
\nc{\bmP}{{\bm P}} \nc{\bmQ}{{\bm Q}} \nc{\bmR}{{\bm R}}
\nc{\bmS}{{\bm S}} \nc{\bmT}{{\bm T}} \nc{\bmU}{{\bm U}}
\nc{\bmV}{{\bm V}} \nc{\bmW}{{\bm W}} \nc{\bmX}{{\bm X}}
\nc{\bmZ}{{\bm Z}}

\nc{\cA}{{\cal A}} \nc{\cB}{{\cal B}} \nc{\cC}{{\cal C}}
\nc{\cD}{{\cal D}} \nc{\cE}{{\cal E}} \nc{\cF}{{\cal F}}
\nc{\cG}{{\cal G}} \nc{\cH}{{\cal H}} \nc{\cI}{{\cal I}}
\nc{\cJ}{{\cal J}} \nc{\cK}{{\cal K}} \nc{\cL}{{\cal L}}
\nc{\cM}{{\cal M}} \nc{\cN}{{\cal N}} \nc{\cO}{{\cal O}}
\nc{\cP}{{\cal P}} \nc{\cQ}{{\cal Q}} \nc{\cR}{{\cal R}}
\nc{\cS}{{\cal S}} \nc{\cT}{{\cal T}} \nc{\cU}{{\cal U}}
\nc{\cV}{{\cal V}} \nc{\cW}{{\cal W}} \nc{\cX}{{\cal X}}
\nc{\cZ}{{\cal Z}}
\begin{document}

\title{$H_2$-reducible matrices in six-dimensional mutually unbiased bases}

\author{Xiaoyu Chen}
\affiliation{LMIB, SKLSDE, BDBC and School of Mathematical Sciences, Beihang University, Beijing 100191, China}

\author{Mengfan Liang}\email[]{sy1809122@buaa.edu.cn (corresponding author)}
\affiliation{LMIB(Beihang University), Ministry of Education, and School of Mathematical Sciences, Beihang University, Beijing 100191, China}

\author{Mengyao Hu}\email[]{mengyaohu@buaa.edu.cn (corresponding author)}
\affiliation{LMIB(Beihang University), Ministry of Education, and School of Mathematical Sciences, Beihang University, Beijing 100191, China}

\author{Lin Chen}\email[]{linchen@buaa.edu.cn (corresponding author)}
\affiliation{LMIB(Beihang University), Ministry of Education, and School of Mathematical Sciences, Beihang University, Beijing 100191, China}
\affiliation{International Research Institute for Multidisciplinary Science, Beihang University, Beijing 100191, China}

\begin{abstract}
Finding four six-dimensional mutually unbiased bases (MUBs) containing the identity matrix is a long-standing open problem in quantum information. We show that if they exist, then the $H_2$-reducible matrix in the four MUBs has exactly nine $2\times2$ Hadamard submatrices. We apply our result to exclude from the four MUBs some known CHMs, such as symmetric $H_2$-reducible matrix, the Hermitian matrix, Dita family, Bjorck's circulant matrix, and Szollosi family. Our results represent the latest progress on the existence of six-dimensional MUBs. 
\end{abstract}

\date{\today}

\maketitle



%\Large

\tableofcontents

\section{Introduction}
\label{sec:intro}

In quantum physics, mutually unbiased bases (MUBs) present a basic notion of describing physical observables \cite{schwinger60}. MUBs have been extensively useful in quantum tomography, discrete Wigner functions
 \cite{Assw14,Assw15} and King problem \cite{mub09}. In particular, MUBs minimize the uncertainty of estimating density matrices and may conceal security in quantum key distribution protocols.
It's been proven that the complete set of $d$-dimensional MUBs has $d+1$ MUBs. The main problem on MUBs is to prove whether MUBs in the $d$-dimensional Hilbert space $\bbC^d$ is complete for any integer $d$. 
It has been proven true when $d$ is prime power. However it is widely conjectured that four MUBs may not exist in $\bbC^6$. Much effort has been devoted to the problem in the past decades \cite{bw08,bw09,bw10,mb15,Sz12,Turek2016A,NICOARA2019143,Sz2008Parametrizing,Goyeneche13,Boykin05,jmm09,deb10,wpz11,mw12ijqi,mw12jpa135307,rle11,mw12jpa102001,mpw16, Chen2017Product,Chen2018Mutually, Designolle2018Quantifying}.

In this paper we investigate this conjecture in terms of the so-called $H_2$-reducible matrix \cite{karlsson11,karlsson11h2}. The latter is a $6\times6$ complex Hadamard matrix (CHM) containing a $2\times2$ Hadamard submatrices, which is proportional to a $2\times2$ unitary matrix. The $H_2$-reducible matrix represents a large subset of CHMs covering many known affine CHMs say the Fourier matrix and non-affine CHMs such as Hermitean family. So the $H_2$-reducible matrix plays an important role whose existence in an MUB trio is worth being studied. 
In \cite{Liang2019}, we have investigated the four-MUB conjecture in terms of the $H_2$-reducible matrix and concluded that the $H_2$-reducible matrix belonging to an MUB trio has exactly nine or eighteen $2\times2$ Hadamard submatrices. In this paper,we concentrate on the problem further.
We review preliminary results on CHMs and $H_2$-reducible matrices in Lemma \ref{le:h2reducible}. Next we review preliminary results on MUB trio in Lemma \ref{le:mubtrio} and \ref{le:linearalg}. 
In Theorem \ref{thm:h2=9only}, we show that if an $H_2$-reducible matrix belongs to an MUB trio, then the matrix has exactly nine $2\times2$ Hadamard submatrices.	This is the main result of this paper supported by the preliminary Lemma \ref{le:h2reducible=9or18=2x2subuni} and \ref{le:h33=-1}. We furthermore apply our result to exclude some known CHMs as members of MUB trio. They include the affine CHMs say the Fourier matrix, Dita family, Bjorck's circulant matrix, and non-affine CHMs such as Hermitean and Szollosi family. Our results present the latest progress on the existence of four six-dimensional MUBs. They are also related to other topics in quantum information, e.g., unitary matrices, tensor rank and unextendible product basis \cite{Baerdemacker2017The,Chen2018The,kai=lma}.

The rest of this paper is structured as follows. In Sec. \ref{sec:pre} we construct the notion of CHMs, equivalence and complex equivalence of matrices, as well as the parametrization of $H_2$-reducible matrices. In Sec. \ref{sec:h2re} we introduce the main result of this paper. In Sec. \ref{sec:app} we apply our result to exclude some known CHMs as members of MUB trio. We conclude in Sec. \ref{sec:con}.

\section{Preliminaries}
	\label{sec:pre}

In this section we introduce the notations and facts used in this paper. We refer to the $n\times n$ complex Hadamard matrix (CHM) $H_n=[u_{ij}]_{i,j=1,...,n}$ as a  matrix with orthogonal row vectors and entries of modulus one. That is,
$
H_n^\dag H_n = nI_n$ and $
\abs{u_{ij}} = 1.
$ To find out the connection between different CHMs, we define the equivalence and complex equivalence. We refer to the monomial unitary matrix as a unitary matrix each of whose row and columns has exactly one nonzero entry, and it has modulus one. Two $n\times n$ matrices $U$ and $V$ are \textit{complex equivalent} when $U=PVQ$ where $P,Q$ are both monomial unitary matrices. If $P,Q$ are both permutation matrices then we say that $U,V$ are \textit{equivalent}. Evidently if $U,V$ are equivalent then they are complex equivalent, and the converse fails. The number of real entries of a CHM may be changed under complex equivalence, while it is unchanged under equivalence. For example, it is straightforward to show that any $n\times n$ CHM is complex equivalent to a CHM containing at least $2n+1$ entry one. They are in the first column and row of the CHM. 

In quantum physics, a pure state is described by a unit vector in linear algebra. Two states in $\bbC^d$ are MU when their inner product is of modulus ${1\over\sqrt d}$. 
Two MUBs are orthonormal basis are MU when their elements are pairwise MU. For convenience we refer to a unitary matrix as an MUB consisting of the column vectors of the unitary matrix. For $d=6$, it has been a long-standing open problem whether four MUBs $I_6,V,W,X$ exist. If it exists then we refer to the three unitary matrices $V,W,X$ as an MUB trio.

In the following we review Theorem 11 of the paper \cite{karlsson11}. We shall use it in the proof of Lemma \ref{le:h33=-1} and Theorem \ref{thm:h2=9only}. The result parameterizes every $H_2$-reducible matrix.
\begin{lemma}
\label{le:h2reducible}
(i) The $H_2$-reducible CHM is complex equivalent to the CHM $H$ in \cite[Theorem 11]{karlsson11}, namely
\bea
\label{eq:h2-1}
H=&&
\bma
F_2 & Z_1 & Z_2\\
Z_3 & {1\over2}Z_3AZ_1 & {1\over2}Z_3BZ_2\\
Z_4 & {1\over2}Z_4BZ_1 & {1\over2}Z_4AZ_2\\
\ema
\notag\\=&&
\bma
I_2 & 0 & 0\\
0 & Z_3 & 0\\
0 & 0 & Z_4\\
\ema
\cdot 
\bma
F_2 & I_2 & I_2\\
I_2 & {1\over2}A & {1\over2}B\\
I_2 & {1\over2}B & {1\over2}A\\
\ema
\cdot
\bma
I_2 & 0 & 0\\
0 & Z_1 & 0\\
0 & 0 & Z_2\\
\ema,
\eea
where 
\begin{eqnarray}
\label{eq:f2z1z2}
&&
F_2=\bma1&1\\1&-1 \ema,	
\quad
Z_1=\bma1&1\\z_1&-z_1 \ema,	
\quad
Z_2=\bma1&1\\z_2&-z_2 \ema,
\notag\\&&	
Z_3=\bma1&z_3\\1&-z_3 \ema,	
\quad
Z_4=\bma1&z_4\\1&-z_4 \ema,	
\notag\\&&
A=
\bma 
A_{11} & A_{12}
\\
A_{12}^* & -A_{11}^*
\ema,
\quad
B=
\bma 
-1-A_{11} & -1-A_{12}
\\
-1-A_{12}^* & 1+A_{11}^*
\ema,
\notag\\&&
A_{11}=-{1\over2}+i{\sqrt3\over2}(\cos\t+e^{-i\phi}\sin\t),
\notag\\&&	
A_{12}=-{1\over2}+i{\sqrt3\over2}(-\cos\t+e^{i\phi}\sin\t),
\notag\\&&
B_{11}=-{1\over2}-i{\sqrt3\over2}(\cos\t+e^{-i\phi}\sin\t),
\notag\\&&	
B_{12}=-{1\over2}-i{\sqrt3\over2}(-\cos\t+e^{i\phi}\sin\t),
\notag\\&&
\t,\phi\in[0,2\pi),
\quad
\abs{z_j}=1,
\notag\\&&
z_3^2=\cM_A(z_1^2)=\cM_B(z_2^2),
\notag\\&&
z_4^2=\cM_A(z_2^2)=\cM_B(z_1^2),
\notag\\&&
\cM_A(z)=
{
A_{12}^2z-A_{11}^2
\over
(A_{11}^2)^*z
-(A_{12}^2)^*
},
\notag\\&&
\cM_B(z)=
{
B_{12}^2z-B_{11}^2
\over
(B_{11}^2)^*z
-(B_{12}^2)^*
}.
\end{eqnarray}

(ii) Suppose $M=
\bma 
F_2 & Z_1 & Z_2\\ 
Z_3 & a & b \\
Z_4 & c & d \\
\ema$ is an $H_2$-reducible matrix where $F_2,Z_1,Z_2,Z_3$ and $Z_4$ are given in \eqref{eq:f2z1z2}. Then $M$ is the same as the matrix $H$ in \eqref{eq:h2-1} satisfying \eqref{eq:f2z1z2}. In particular, $a,b,c,d$ are $2\times2$ Hadamard submatrices described in \eqref{eq:f2z1z2}.
\qed	
\end{lemma}

We review Theorem 12 in the recent paper \cite{Liang2019} on $H_2$-reducible matrices and MUBs. This is the main result of \cite{Liang2019}. We shall use it in the proof of Theorem \ref{thm:h2=9only} as the main result of this paper.
\begin{lemma}
\label{le:h2reducible=9or18=2x2subuni}
If an $H_2$-reducible matrix belongs to an MUB trio, then the matrix has exactly nine or eighteen $2\times2$ Hadamard submatrices. 
\qed
\end{lemma}

Next we review a fact from \cite[Lemma 11]{Chen2017Product}. It gives the necessary condition by which a $6\times6$ CHM is a member of some MUB trio. This is used in the proof of Lemma \ref{le:h33=-1}. 

\begin{lemma}
\label{le:mubtrio}
The CHM in an MUB trio contains no a real $2\times3$ or $3\times2$ submatrix up to complex equivalence.
\end{lemma}

Finally we review a fact on complex numbers used in the proof of Theorem \ref{thm:h2=9only}.
\begin{lemma}
\label{le:linearalg}	
(i) Suppose $a+b+c=0$ with complex numbers $a,b,c$ of modulus one. Then $(a,b,c)\propto(1,\o,\o^2)$ or $(1,\o^2,\o)$ with $\o=e^{2\p i\over 3}$.

(ii) Suppose $a+b+c+d=0$ with complex numbers $a,b,c,d$ of modulus one. Then $a=-b,-c$ or $-d$.
\end{lemma}

\section{The $H_2$-reducible matrix in an MUB trio}
\label{sec:h2re}

In this section we show that the $H_2$-reducible matrix in an MUB trio has exactly nine $2\times2$ Hadamard submatrices. This is presented in Theorem \ref{thm:h2=9only}. For this purpose we construct a preliminary lemma.
\begin{lemma}
\label{le:h33=-1}
Suppose $M$ is a $6\times6$ CHM containing more than nine $2\times2$ Hadamard submatrices, and $M$ belongs to an MUB trio. Then up to complex equivalence we may assume that $M$ is the matrix in \eqref{eq:h2-1} with the entry $(3,3)$ of $M$ being $-1$.
\end{lemma}
\begin{proof}
Evidently $M$ is an $H_2$-reducible matrix. It follows from Lemma \ref{le:h2reducible} (i) that there exist two $6\times6$ monomial unitary matrices $P,Q$ such that
\bea
\label{eq:h2-0-mengfan}
H:=&& [h_{ij}]:=H(\t,\phi,z_1,z_2,z_3,z_4)=PMQ
\notag\\
=&&
\bma
F_2 & Z_1 & Z_2\\
Z_3 & {1\over2}Z_3AZ_1 & {1\over2}Z_3BZ_2\\
Z_4 & {1\over2}Z_4BZ_1 & {1\over2}Z_4AZ_2\\
\ema
\notag\\
=&&
\bma
I_2 & 0 & 0\\
0 & Z_3 & 0\\
0 & 0 & Z_4\\
\ema
\cdot 
\bma
F_2 & I_2 & I_2\\
I_2 & {1\over2}A & {1\over2}B\\
I_2 & {1\over2}B & {1\over2}A\\
\ema
\cdot
\bma
I_2 & 0 & 0\\
0 & Z_1 & 0\\
0 & 0 & Z_2\\
\ema,
\eea
where $F_2,Z_1,...,Z_4,A,B$ containing the parameters $\t,\phi,z_1,z_2,z_3,z_4$ with $\t,\phi\in[0,\pi),$ and $
\abs{z_j}=1$ are given in \eqref{eq:f2z1z2}. Since $H$ and $M$ are complex equivalent, $H$ still has more than nine $2\times2$ Hadamard submatrices. Using \eqref{eq:h2-0-mengfan}, we obtain that $h_{ij}=-1$ for some $(i,j)$ such that $i,j\ne1$ and $(i,j)\ne(2,2)$. If one of $z_1,...,z_4$ is $-1$ then $H$ has a $2\times4$ or $4\times2$ real submatrix. It is a contradiction with Lemma \ref{le:mubtrio}, because $H$ belongs to an MUB trio. Hence $h_{ij}=-1$ with $(i,j)\in\{3,4,5,6\}\times\{3,4,5,6\}$.

We present claim one as follows. If we assume that $H$ in \eqref{eq:h2-0-mengfan} with $h_{33},h_{35},h_{53}$ or $h_{55}=-1$ does not belong to any MUB trio, then neither does $H$ with $h_{ij}=-1$ and $(i,j)\in\{3,4,5,6\}\times\{3,4,5,6\}$. To prove the claim, we consider $H(\t,\phi,z_1,z_2,z_3,z_4)$ with $h_{43}=-1$. Let the permutation matrix $R=I_2\oplus\bma0&1\\1&0 \ema\oplus I_2$. Then \eqref{eq:h2-0-mengfan} implies that $RH(\t,\phi,z_1,z_2,z_3,z_4)=H(\t,\phi,z_1,z_2,-z_3,z_4)=[h'_{ij}]$ with $h'_{33}=-1.$ Using the assumption, we obtain that $H(\t,\phi,z_1,z_2,-z_3,z_4)=[h'_{ij}]$ does not belong to any MUB trio. Neither does $H(\t,\phi,z_1,z_2,z_3,z_4)$ with $h_{43}=-1$. One can similarly show that $H(\t,\phi,z_1,z_2,z_3,z_4)$ with $h_{34},h_{44},...$ or $h_{66}=-1$. We have proven claim one. 

We present claim two as follows. If we assume that $H$ in \eqref{eq:h2-0-mengfan} with $h_{33}=-1$ does not belong to any MUB trio, then neither does $H$ with $h_{35},h_{53},h_{55}=-1$. To prove the claim, we consider $H(\t,\phi,z_1,z_2,z_3,z_4)$ with $h_{55}=-1$. Let the permutation matrix $R_1=I_2\oplus
\bma 
0&0&1&0\\ 
0&0&0&1\\ 
1&0&0&0\\ 
0&1&0&0\\ 
\ema$. Then \eqref{eq:h2-0-mengfan} implies that $R_1H(\t,\phi,z_1,z_2,z_3,z_4)R_1=H(\t,\phi,z_2,z_1,z_4,z_3)=[h''_{ij}]$ with $h''_{33}=-1.$ Using the assumption, we obtain that $[h''_{ij}]$ does not belong to any MUB trio. We have proven claim two. One can similarly show that $H$ with $h_{35}=-1$ belongs to an MUB trio if and only if so does $H$ with $h_{53}=-1$. To prove the claim, if suffices to show that $H(\t,\phi,z_1,z_2,z_3,z_4)$ with $h_{53}=-1$ does not belong to any MUB trio. Indeed, \eqref{eq:h2-0-mengfan} implies that $R_1H(\t,\phi,z_1,z_2,z_3,z_4)=H(\t+\p,\phi,z_1,z_2,z_4,z_3)=[h'''_{ij}]$ with $h'''_{33}=-1.$ Using the assumption, we obtain that $[h'''_{ij}]$ does not belong to any MUB trio. We have proven claim two.

By combining claim one and two, we have proven the assertion. 
\end{proof}

From the proof of Lemma \ref{le:h33=-1}, one can similarly obtain the following observation.
\begin{corollary}
\label{le:h33=x}
Suppose $M$ is an $H_2$-reducible matrix in \eqref{eq:h2-1}. If $M$ with $h_{33}=x$ does not belong to any MUB trio, then neither does $M$ with $h_{ij}=x$ and $(i,j)$ is one of $(3,4),(3,5),...,(6,6)$. 
\end{corollary}

Now we are in a position to prove the main result of this section. 

\begin{theorem}
\label{thm:h2=9only}
If an $H_2$-reducible matrix belongs to an MUB trio, then the matrix has exactly nine $2\times2$ Hadamard submatrices.	

In other words, the member of MUB trio has no CHM containing the $3\times3$ submatrix
$\bma
1&1&1\\
1&-1&*\\
1&*&-1\\
\ema$ up to complex equivalence.
\end{theorem}
\begin{proof}
Suppose $M=[m_{ij}]$ is an $H_2$-reducible matrix belonging to an MUB trio. It follows from Lemma \ref{le:h2reducible=9or18=2x2subuni} that $M$ has exactly nine or eighteen $2\times2$ Hadamard submatrices. We shall exclude the latter by contradiction, and the assertion follows. 

Assume that $M$ has exactly eighteen $2\times2$ Hadamard submatrices. Using Lemma \ref{le:h33=-1}, we may assume that $M$ is the matrix in \eqref{eq:h2-1} with $m_{33}=-1$. Applying Lemma \ref{le:linearalg} (ii) to row $1,3$ of $M$ and column $5,6$ of $M$, we obtain that one of $\bma m_{12} & m_{15} \\ m_{32} & m_{35} \ema$ and $\bma m_{12} & m_{16} \\ m_{32} & m_{36} \ema$ is an Hadamard submatrix. Let the permutation matrix $R_1=I_2\oplus I_2\oplus\bma0&1\\1&0 \ema$. Using \eqref{eq:h2-1}, one can show that $MR_1$ is still an $H_2$-reducible matrix containing exactly eighteen $2\times2$ Hadamard submatrices in an MUB trio. For convenience we still name $MR_1$ as $M=[m_{ij}]$. So $\bma m_{12} & m_{15} \\ m_{32} & m_{35} \ema$ is a $2\times2$ Hadamard submatrix. Similarly by studying column $1,3$, we may assume that $\bma m_{21} & m_{23} \\ m_{51} & m_{53} \ema$ is a $2\times2$ Hadamard submatrix. Using Lemma \ref{le:h2reducible}, we can determine the eighteen $2\times2$ Hadamard submatrices in $M$. We obtain four equations $m_{33}=-1$, $m_{35}=-z_3$, $m_{53}=-z_1$, and $m_{55}=z_2z_4$. Using \eqref{eq:h2-1}, one can derive the expressions of $M=[m_{ij}]$. By solving the four equations, we obtain $m_{66}=-1$. It means that $M$ has more than eighteen $2\times2$ Hadamard submatrices. It is a contradiction with the assumption that $M$ has exactly eighteen $2\times2$ Hadamard Hadamard submatrices. So we have excluded the option that $M$ has exactly eighteen $2\times2$ Hadamard submatrices. We have proven the assertion. 
\end{proof}

In the next section we introduce the application of Theorem \ref{thm:h2=9only}.

\section{application}
\label{sec:app}

In this section, we will exclude some known CHMs as members of MUB trio by using Theorem \ref{thm:h2=9only}. First of all we introduce Theorem \ref{le:shrminchm} and Theorem \ref{le:Hermitian matrix}, which exculde symmetric $H_2$-reducible matrix and the Hermitian matrix respectively from MUB trio.
\begin{theorem}
\label{le:shrminchm}	
The CHM in any MUB trio is not a symmetric $H_2$-reducible matrix.

\end{theorem}
\begin{proof}
Suppose that $H$ is a symmetric $H_2$-reducible matrix in an MUB trio. The first row of $H$ is $(h_1,h_2,h_3,h_4,h_5,h_6)$. Let $D=\diag(h_1^{-\frac{1}{2}},h_1^{-\frac{1}{2}}h_2^{-1},h_1^{-\frac{1}{2}}h_3^{-1},h_1^{-\frac{1}{2}}h_4^{-1},h_1^{-\frac{1}{2}}h_5^{-1},h_1^{-\frac{1}{2}}h_6^{-1})$. Then $D$ is a monomial unitary matrix, and $H^{'}=DHD$ is a symmetric $H_2$-reducible matrix whose first row and column consist of ones. According to the Corollary 3 of \cite{karlsson11}, $H^{'}$ has at least one element equaling to $-1$. If the element $-1$ does not belong to the diagonal of $H^{'}$, then $H^{'}$ contains at least two elements equaling to $-1$ by the symmetry of $H^{'}$. Hence $H^{'}$ does not belong to any MUB trio by Theorem \ref{thm:h2=9only}. So the element $-1$ belongs to the diagonal of $H^{'}$. From the proof of Lemma \ref{le:h33=-1}, we know that there is a permutation matrix $P$ s.t. $H^{''}=PH^{'}P$, and the element of the third row and the third column of $H^{''}$ is $-1$, meanwhile $H^{''}$ is a symmetric $H_2$-reducible matrix which has the form in \eqref{eq:h2-1}. We assume that
\bea
H^{''}=&&
\bma
F_2 & Z_1 & Z_2\\
Z_3 & {1\over2}Z_3AZ_1 & {1\over2}Z_3BZ_2\\
Z_4 & {1\over2}Z_4BZ_1 & {1\over2}Z_4AZ_2\\
\ema
\notag
\eea
where 
\begin{eqnarray}
\label{eq:1f2z1z2}
&&
F_2=\bma1&1\\1&-1 \ema,	
\quad
Z_1=\bma1&1\\z_1&-z_1 \ema,	
\quad
Z_2=\bma1&1\\z_2&-z_2 \ema,
\notag\\&&	
Z_3=\bma1&z_3\\1&-z_3 \ema,	
\quad
Z_4=\bma1&z_4\\1&-z_4 \ema,	
\notag\\&&
A=
\bma 
A_{11} & A_{12}
\\
A_{12}^* & -A_{11}^*
\ema,
\quad
B=
\bma 
-1-A_{11} & -1-A_{12}
\\
-1-A_{12}^* & 1+A_{11}^*
\ema,
\notag\\&&
A_{11}=-{1\over2}+i{\sqrt3\over2}(\cos\t+e^{-i\phi}\sin\t),
\notag\\&&	
A_{12}=-{1\over2}+i{\sqrt3\over2}(-\cos\t+e^{i\phi}\sin\t),
\notag\\&&
B_{11}=-{1\over2}-i{\sqrt3\over2}(\cos\t+e^{-i\phi}\sin\t),
\notag\\&&	
B_{12}=-{1\over2}-i{\sqrt3\over2}(-\cos\t+e^{i\phi}\sin\t),
\notag\\&&
\t,\phi\in[0,2\pi),
\quad
\abs{z_j}=1,
\notag\\&&
z_3^2=\cM_A(z_1^2)=\cM_B(z_2^2),
\notag\\&&
z_4^2=\cM_A(z_2^2)=\cM_B(z_1^2),
\notag\\&&
\cM_A(z)=
{
A_{12}^2z-A_{11}^2
\over
(A_{11}^2)^*z
-(A_{12}^2)^*
},
\notag\\&&
\cM_B(z)=
{
B_{12}^2z-B_{11}^2
\over
(B_{11}^2)^*z
-(B_{12}^2)^*
}.
\end{eqnarray}
Suppose that $h_{mn}^{''}(1\leq m,n\leq 6)$ is the element of the $m^{'}$th row and the $n^{'}$ column of $H^{''}$. Because of the symmetry of $H^{''}$, we have $h_{mn}^{''}=h_{nm}^{''}$, $z_1=z_3$ and $z_2=z_4$. By $h_{34}^{''}=h_{43}^{''}$,
we have 
\begin{eqnarray}
\frac{1}{2}(A_{11}+A_{12}z_1-\overline{A_{12}}z_3+\overline{A_{11}}z_1z_3)=\frac{1}{2}(A_{11}-A_{12}z_1+z_3(\overline{A_{12}}+\overline{A_{11}}z_1)).
\end{eqnarray}
Namely
\begin{eqnarray}
(-\cos\t+e^{i\phi}\sin\t)z_1=0.
\end{eqnarray}
Obviously $z_1\neq 0$. So $-\cos\t+e^{i\phi}\sin\t = 0$, namely $\tan\t =e^{-i\phi}$. Since $\tan\t$ is real, then $e^{-i\phi}=1$ or $e^{-i\phi}=-1$. Now we can verify that $h_{45}^{''}=h_{54}^{''}=-1$ by using \eqref{eq:1f2z1z2}. Hence $H^{''}$ dose not belong to any MUB trio by Theorem \ref{thm:h2=9only}. To sum up we have $H$ dose not belong to any MUB trio. So we complete this proof.

\end{proof}

\begin{theorem}
\label{le:Hermitian matrix}	
The CHM in any MUB trio is not an Hermitian matrix.
\end{theorem}
\begin{proof}
Suppose that $H_6$ is an Hermitian matrix in an MUB trio. From the paper \cite{HTCHM} by Kyle Beauchamp and Remus Nicoara we know that $H_6$ is equivalent to $H(\t)$, where:
\begin{eqnarray}
H(\t)=
\bma
1& 1 & 1 & 1 & 1 & 1 \\
1&-1 & \frac{1}{x} & -y & -\frac{1}{x} & y \\
1& x & -1 & t & -t & -x \\
1& -\frac{1}{y}& \frac{1}{t} & -1 & \frac{1}{y} & -\frac{1}{t} \\
1& -x & -\frac{1}{t} & y & 1 & \frac{1}{z}\\
1& \frac{1}{y} & -\frac{1}{x} & -t & z & 1 \\
\ema,
\end{eqnarray}
and $\t \in [-\pi ,-arcos (\frac{-1+\sqrt{3}}{2})]\cup [arcos (\frac{-1+\sqrt{3}}{2}),\pi ]$, the parameters $x,y,z,t$ are given by:
\begin{eqnarray}
&y=e^{i\t}, z=\dfrac{1+2y-y^2}{y(-1+2y+y^2)}\\
&x=\dfrac{1+2y+y^2-\sqrt{2}\sqrt{1+2y+2y^3+y^4}}{1+2y-y^2}\\
&t=\dfrac{1+2y+y^2-\sqrt{2}\sqrt{1+2y+2y^3+y^4}}{-1+2y+y^2}.
\end{eqnarray}
One can verify that $H(\t)$ is an $H_2$-reducible matrix of more than eighteen $2\times 2$ Hadamard submatrices. So $H(\t)$ does not belong to any MUB trio by Theorem \ref{thm:h2=9only}, of cause $H_6$ does not belong to any MUB trio. Hence we complete this proof.
\end{proof}

Next we shall investigate some affine and non-affine CHMs. First the two CHMs constructed on p256 of \cite{karlsson11}, and the Dita family \cite[Eq. (5.45)]{deb10} which are affine CHMs, all have more than nine $2\times2$ Hadamard submatrices. So they are both excluded by Theorem \ref{thm:h2=9only}. One can similarly exclude the Fourier family and its transpose as known affine families. Next a special CHM is Bjorck's circulant matrix \cite[Eq. (5.46)]{deb10}, 
\begin{eqnarray}
C_6=
\bma
1 & id & -d & -i & -d^* & id^* \\
id^*& 1 & id & -d & -i & -d^* \\
-d^* & id^* & 1 & id & -d & -i \\
-i & -d^* & id^* & 1 & id & -d \\
-d & -i & -d^* & id^* & 1 & id\\
id & -d & -i & -d^* & id^* & 1 \\
\ema,
\end{eqnarray}
where $d={1-\sqrt3\over2}+i\sqrt{{\sqrt3\over2}}$. One can show that $C_6$ has more than nine $2\times2$ Hadamard submatrices. 
It is known that every circulant Hadamard matrix is equivalent to either the $6\times6$ Fourier matrix or $C_6$. So every circulant Hadamard matrix does not belong to any MUB trio.

Third Theorem \ref{thm:h2=9only} excludes some non-affine CHMs too, such as the Szollosi family \cite[Eq. (C.12)]{mub09}
\begin{eqnarray}
X(a,b)=H(x,y,u,v)=
\bma
1& 1& 1& 1& 1& 1\\
1& x^2y& xy^2& {xy\over uv}& uxy& vxy\\
1& {x\over y}& x^2y& {x\over u} & {x\over v}& uvx \\
1& uvx& uxy& -1& -uxy& -uvx\\
1& {x\over u}& vxy& -{x\over u}& -1& -vxy\\
1& {x\over v} & {xy\over uv} & -{xy\over uv}& -{x\over v} & -1\\
\ema,
\end{eqnarray}
where entries $x,y$ and $u,v$ are solutions of the equations $f_{\a}=0$ and $f_{-\a}=0$, respectively such that $f_{\a}(z)=z^3-\a z^2+\a^*z-1$ and $\a=a+bi$ restricted by $D(\a)\le0$ and $D(-\a)\le0$ with $D(\a)=\abs{\a}^4+18\abs{\a}^2-8 Re[\a^3]-27$. One can show that both Hermitean and Szollosi families are $H_2$-reducible matrices of more than nine $2\times2$ Hadamard submatrices. So they are not members of any MUB trio in terms of Theorem \ref{thm:h2=9only}.

\section{Conclusions}
\label{sec:con}

We have shown that if four six-dimensional MUBs containing the identity matrix exist, then the $H_2$-reducible matrix in the four MUBs has exactly nine $2\times2$ Hadamard submatrices. We have applied our result to exclude some known affine and non-affine CHMs as members of MUB trio, such as symmetric $H_2$-reducible matrix, the Hermitian matrix, Dita family, Bjorck's circulant matrix, and Szollosi family. The next step is to exclude every $H_2$-reducible matrix as a member of MUB trio.

\section*{Acknowledgements}

XYC and MFL are supported by NSFC (Grant No. 61702025) and State Key Laboratory of Software Development Environment  (Grant No. SKLSDE-2019ZX-12). MYH and LC were supported by the  NNSF of China (Grant No. 11871089), and the Fundamental Research Funds for the Central Universities (Grant No. ZG216S1902).

\bibliographystyle{unsrt}

\bibliography{mengfan=6x6real20}

\end{document}